\documentstyle{mn}

\newif\ifAMStwofonts

\ifoldfss
  \ifCUPmtlplainloaded \else
    \NewTextAlphabet{textbfit} {cmbxti10} {}
    \NewTextAlphabet{textbfss} {cmssbx10} {}
    \NewMathAlphabet{mathbfit} {cmbxti10} {} 
    \NewMathAlphabet{mathbfss} {cmssbx10} {} 
  \fi
  \ifAMStwofonts
    \ifCUPmtlplainloaded \else
      \NewSymbolFont{upmath} {eurm10}
      \NewSymbolFont{AMSa} {msam10}
      \NewMathSymbol{\upi}     {0}{upmath}{19}
      \NewMathSymbol{\umu}     {0}{upmath}{16}
      \NewMathSymbol{\upartial}{0}{upmath}{40}
      \NewMathSymbol{\leqslant}{3}{AMSa}{36}
      \NewMathSymbol{\geqslant}{3}{AMSa}{3E}

    \fi
  \fi
\fi 

\ifnfssone
  \newmathalphabet{\mathit}
  \addtoversion{normal}{\mathit}{cmr}{m}{it}
  \addtoversion{bold}{\mathit}{cmr}{bx}{it}
  \newmathalphabet{\mathbfit} 
  \addtoversion{normal}{\mathbfit}{cmr}{bx}{it}
  \addtoversion{bold}{\mathbfit}{cmr}{bx}{it}
  \newmathalphabet{\mathbfss} 
  \addtoversion{normal}{\mathbfss}{cmss}{bx}{n}
  \addtoversion{bold}{\mathbfss}{cmss}{bx}{n}
  \ifAMStwofonts
    \ifCUPmtlplainloaded \else
      \UseAMStwoboldmath
      \makeatletter
      \new@mathgroup\upmath@group
      \define@mathgroup\mv@normal\upmath@group{eur}{m}{n}
      \define@mathgroup\mv@bold\upmath@group{eur}{b}{n}
      \edef\UPM{\hexnumber\upmath@group}
      \new@mathgroup\amsa@group
      \define@mathgroup\mv@normal\amsa@group{msa}{m}{n}
      \define@mathgroup\mv@bold\amsa@group{msa}{m}{n}
      \edef\AMSa{\hexnumber\amsa@group}
      \makeatother
      \mathchardef\upi="0\UPM19
      \mathchardef\umu="0\UPM16
      \mathchardef\upartial="0\UPM40
      \mathchardef\leqslant="3\AMSa36
      \mathchardef\geqslant="3\AMSa3E
    \fi
  \fi
\fi 

\ifnfsstwo
  \DeclareMathAlphabet{\mathbfit}{OT1}{cmr}{bx}{it}
  \SetMathAlphabet\mathbfit{bold}{OT1}{cmr}{bx}{it}
  \DeclareMathAlphabet{\mathbfss}{OT1}{cmss}{bx}{n}
  \SetMathAlphabet\mathbfss{bold}{OT1}{cmss}{bx}{n}
  \ifAMStwofonts
    \ifCUPmtlplainloaded \else
      \DeclareSymbolFont{UPM}{U}{eur}{m}{n}
      \SetSymbolFont{UPM}{bold}{U}{eur}{b}{n}
      \DeclareSymbolFont{AMSa}{U}{msa}{m}{n}
      \DeclareMathSymbol{\upi}{0}{UPM}{"19}
      \DeclareMathSymbol{\umu}{0}{UPM}{"16}
      \DeclareMathSymbol{\upartial}{0}{UPM}{"40}
      \DeclareMathSymbol{\leqslant}{3}{AMSa}{"36}
      \DeclareMathSymbol{\geqslant}{3}{AMSa}{"3E}
    \fi
  \fi
\fi 

\ifCUPmtlplainloaded \else
  \ifAMStwofonts \else 
    \def\upi{\pi}
    \def\umu{\mu}
    \def\upartial{\partial}
  \fi
\fi

\title{Global Textures and the Formation\\ of Self-Bound
Gravitational Systems}
\author[A. L. B. Ribeiro and P. S. Letelier]
       {A. L. B. Ribeiro and P. S. Letelier\\
        Departamento de Matem\'atica Aplicada - Instituto de
Matem\'atica Estat\'{\i}stica e Computa\c c\~ao Cient\'{\i}fica,\\
Universidade Estadual de Campinas, 13083-970 SP, Brazil}
\date{Accepted 1999.
      Received 1999;
      in original form 1999 February 15}

\pagerange{\pageref{firstpage}--\pageref{lastpage}}
\pubyear{1998}

\begin{document}

\maketitle

\label{firstpage}

\begin{abstract}
Using a Newtonian approximation we developed
a quantitative criterion for the collapse of 
a spherical distribution of matter
under an isolated texture field. In particular, we found the
evolution of an overdense region is strongly determined by 
two parameters: the
energy scale of symmetry breaking ($\eta$) and the initial radius of the system. 
Applying our collapse criterion to typical galaxy scales
we verified the formation of $10^{11}M_\odot$ objects 
at $z\la 9$ and $10^{12}M_\odot$ objects at $z\la 5$.

\end{abstract}


\section{Introduction}

The emergence of structures in the universe is one of
the most important problems of modern cosmology. The difficulty
is to understand how an initially homogeneous mass distribution
evolved into its present clumpy state. Generally, it is
accepted that structures were initiated by small density fluctuations
formed at the early universe, and that the subsequent clustering was
produced by gravitational instability (e.g. Kolb \& Turner 1990). 
At the moment, there are two theoretical
approaches proposed to explain the 
origin of the primordial inhomogeneities.
In the context of inflationary models,  they result
of quantum-induced Gaussian fluctuations amplified  over an accelerated
phase of cosmic expansion. On the contrary, 
in topological defect scenarios, non-Gaussian fluctuations are
produced  during
spontaneous symmetry breaking at phase transitions in the early
universe (e.g. Vilenkin \& Shellard 1994). 

Several types of topological defects can be formed depending 
on the kind of symmetry that is broken. Here, we are interested
in a specific topological defect formed during
the symmetry breaking of non-Abelian groups: a
global texture. This defect has been
proposed as a possible source of galaxy formation in 
the universe (Turok 1989).
The development of structures based on  this
topological remnant is related to
the evolution of textures knots.
On scales larger than horizon,
textures knots are regions where the scalar field winds
around the vacuum manifold in a non-trivial way.
As knots come inside the horizon, they collapse at the speed
of light down to an infinitesimal scale where they  unwind themselves emitting  spherical waves of outgoing radiation. This collapse produces an
overdense region onto which matter is
attracted, possibly leading to the formation of non-linear
objects (Turok \& Spergel 1990).

The most desirable way to study structure formation 
produced by textures knots is through
numerical simulations like those carried out by Park, Spergel
\& Turok (1989),
Spergel et al. (1991) and Cen et al. (1991). All these studies indicate
important effects of textures on the distribution of matter, like coherent large-scale flows and 
clustering of galaxies fairly consistent with the correlation function observed
in the nearby universe.
However, we should keep in mind that
the dynamics of textures is highly non-linear since 
the decoupling era (Albrecht, Battye \& Robinson 1997), and that
it is not simple to trace the evolution of the
fluctuations over this long time interval. Thus, all
these numerical works are strongly dependent on  the choice of
the initial conditions for the N-body simulations
(Deruelle et al. 1997). 

Actually, this is not the only problem with the texture scenario.
The work of Pen, Seljak \& Turok (1997) shows
that the texture model
has strongly suppressed acoustic peaks in comparison to 
current observations of CMB anisotropies.
Also, when COBE normalized, the model produces 
a low value of the rms density variation on scales
of 8 Mpc ($\sigma_8 \sim 0.23$), while a
higher value of this parameter (closer to unity) is necessary
in order to get agreement with the observed galaxy clustering. 
These results clearly pose serious difficulties to the
texture scenario. However, this does not mean that the model should be 
completely discarded at the present time,
since a decisive, model discriminating test will be possible
only after the analysis of  the high resolution CMB maps which will
be available in the next years from the MAP and PLANCK satellite
missions. Hence, for the moment, it remains valid to work with
texture knots as a viable way to form galaxies in the universe.

Indeed, an alternative way to study structure formation via global
textures is to develop simple analytic models to describe the general 
properties of the defect dynamics. For instance, Gooding, Spergel
\& Turok (1991)
studied the formation of bound objects induced 
by a number of collapsing knots
in a $\Omega_0=1$ CDM-dominated universe.
An important result of this study is  the prediction
of early spheroidal formation in the universe, suggesting that
by $z\sim 50$, $\sim$3\% of the mass of the universe has formed
non-linear objects with mass greater than $10^6M_\odot$ and that
most objects larger than $10^{12}M_\odot$ formed by $z\sim 2-3$.

In the present work, we use a Newtonian 
approximation to investigate the formation of
self-bound gravitational systems seeded by   
global textures. However, instead of applying a statistical approach to
the contribution of many individual knots, we are only interested
in the local influence of a single texture knot on the process of
galaxy formation.
Our aim is just to obtain a quantitative criterion for the collapse
of a spherical distribution of matter under a texture field.

\section[]{The Collapse Criterion}

Distributed field gradients induced by a number of global
textures are supposed to give a larger contribution
to density inhomogeneities than that produced by
single isolated knots (e.g. Vilenkin \& Shellard 1994).
If this is correct, then, the effort to
probe the process of structure formation seeded
by these defects will require large dynamic range 
simulations in computers of latest technology.
However, the demand for such numerical works does
not mean that analytical developments are unnecessary.
On the contrary, they are particularly useful to
give a supplementary and intuitive picture of the 
defect dynamical contribution 
on structure formation. Specifically, in this work, we
study the process of accretion of matter
onto a single texture knot. This problem has been
studied by several authors over the years. But now,
our emphasis is on
the Newtonian treatment and the possibility of
taking a region where the assumption of an isolated
texture is reasonable.

In order to do that, let's consider
an initially spherical and homogeneous distribution of matter
with density $\rho_b$ when the texture field is turned on. 
The Newtonian density associated to the texture can be
obtained by using the Einstein equations in the weak field
approximation. By numerically solving the Barriola-Vaschaspati
equations for a self-gravitating texture, Gu\'eron \& Letelier (1997)
found that the weak field condition is fulfilled and the
corresponding Newtonian density is

\begin{equation}
\rho_{\rm N} = {8\eta^2(r^2 - t^2)\over(r^2 + t^2)^2}
\end{equation}

\noindent where $\eta$ is the energy scale of symmetry
breaking.  This density is negligible
except when  the texture size, $t$, is small compared
to the physical scale $r$. At the same time, the region where
is reasonable to take account a single texture
configuration, ignoring the effects from other textures,
is limitted by

\begin{equation}
(\Delta t^2 + r^2)^{1/2}\la H^{-1}
\end{equation}

\noindent (N\"otzold 1991), where $\Delta t = t_\ast -t$, $t_\ast$ is
the time at which the texture is collapsed and $H$ is the cosmic
expansion rate.
For simplicity, we will only study the behaviour of regions where  $r\gg|t|$, so we can use an effective  density defined as
\begin{equation}
\rho_{\rm T} \equiv {8\eta^2/ r^2}.
\end{equation}

\noindent For such regions, $|\Delta t| < r$ and so
condition (2) is simply $r\la H^{-1}/\sqrt{2}$.
The total density within r is
$\rho(r)=\rho_{\rm T} + \rho_{\rm b}$ and the density contrast can be
expressed by

\begin{equation}
{\delta\rho\over \rho} = {\rho(r) - \rho_b\over \rho_{\rm b}} = {\rho_{\rm T}\over \rho_{\rm b}}
\end{equation} 

\noindent so that the motion of a thin shell of particles located at r is
governed by the equation

\begin{equation}
{d^2r\over dt^2} = -{GM\over r^2},~~{\rm where}~~
M = {4\pi r^3\over 3}\rho_{\rm b}(1 + \overline{\delta})
\end{equation}

\noindent and the average density contrast is

\begin{equation}
\overline{\delta} = {3\over 4\pi r^3}\int_0^r \left({\rho_{\rm T}\over\rho_{\rm b}}\right)4\pi r^2 \; dr.
\end{equation}

As usual, it will be assumed here 
that the mass within the shell is constant.
In fact, this is strictly correct only after
a time interval of $\sim t_\ast$, when the texture mass
is converted in massless goldstone bosons, leaving behind
the other material components: baryonic and dark matter.
Thus, we are tacitly assuming that $t_\ast$ is much
smaller than the dynamical time of the system, 
but large enough to allow the texture mass
to dominate the first steps of the gravitational instability.
In idealizing the process this way, we hope to make it
well-posed mathematically, although a justification of the
idealization will be only possible from the agreement
(or not) of our predictions with the data.

Now, supposing that the peculiar velocities at $t=t_{\rm i}$ are negligible,
the first integral of Eq.(5) will give the total energy of
the shell and can be written as

\begin{equation}
E = K_{\rm i}\Omega_{\rm i}\lbrack\Omega_{\rm i}^{-1} - (1 + \overline{\delta}_{\rm i})\rbrack,
\end{equation}

\noindent (e.g. Padmanabhan 1993). The subscript $i$ indicates
the initial time $t_{\rm i}$ when the radius of the shell is $r_{\rm i}$,
the kinetic energy is $K_{\rm i}$ and the density parameter is $\Omega_{\rm i}$.
Naturally, the condition $E < 0$ for a self-bound system is

\begin{equation}
\overline{\delta}_{\rm i} > (\Omega_{\rm i}^{-1} -1) \equiv \delta_{\rm c}.
\end{equation}

\noindent For a dust configuration ($p=0$), we have

\begin{equation}
\Omega_{\rm i} = {\Omega_0(1 + z_{\rm i})\over 1 + z_{\rm i}\Omega_0}
\end{equation}

\noindent and so

\begin{equation}
\delta_{\rm c} = {1 - \Omega_0\over\Omega_0(1 +z_{\rm i})}
\end{equation}

\noindent (e.g. B\"orner 1988). Therefore, for the case of a flat
universe, $\delta_{\rm c}=0$ and any overdense
region with $\overline{\delta}_{\rm i} > 0$ will collapse.  
On the other hand, for the case of
an open universe, only regions where $\overline{\delta}_{\rm i} > \delta_{\rm c}$ 
can produce self-bound gravitational systems. In this case,
introducing definition (3) in Eq.(6) we directly find

\begin{equation}
\overline{\delta}_{\rm i} = {24\eta^2\over \rho_{\rm b}(t_{\rm i})r_{\rm i}^2}. 
\end{equation}

Now, it is easy to foresee the destiny of a spherical distribution
of matter under a texture field when $\Omega_0 < 1$. 
Introducing Eq.(11) in condition (8),
it is simple to show that

\begin{equation}
r_{\rm i} < \sqrt{{24\eta^2\Omega_{\rm i}\over \rho_{\rm b}(t_{\rm i})
(1-\Omega_i)}}.
\end{equation}

\noindent Recalling that

\begin{equation}
\rho_{\rm b}(t_{\rm i}) = \Omega_{\rm i}\rho_{\rm c}(t_{\rm i})~~~~~~{\rm and}~~~~~~
\rho_{\rm c}(t_{\rm i}) = {3H_{\rm i}^2\over 8\pi G},
\end{equation}

\noindent we reach

\begin{equation}
r_{\rm i} < {8\eta\over H_{\rm i}}\sqrt{G\pi\over 1 - \Omega_{\rm i}}\equiv R_{\rm c},
\end{equation}

\noindent where $R_{\rm c}$ is the critical radius for the collapse. 
Once more, assuming a Friedmann model with
vanishing pressure and with no cosmological constant 

\begin{equation}
H_{\rm i}^2 = H_0^2\lbrack \Omega_0(1+z_{\rm i})^3+(1-\Omega_0)(1+z_{\rm i})^2\rbrack
\end{equation}

\noindent (e.g. Sandage 1961), we can finally express the collapse
criterion for any overdense region under a texture
field at any epoch $z_{\rm i}$:

\begin{equation}
r_{\rm i} < {8\eta\over H_0}{\sqrt{G\pi(1+z_{\rm i}\Omega_0)\over (1 - \Omega_0)\lbrack \Omega_0(1+z_{\rm i})^3+(1-\Omega_0)(1+z_{\rm i})^2\rbrack}}.
\end{equation}

\section{The Choice of the Energy Scale}

From Eq.(16) we see that, assuming specific values for $H_0~{\rm and}~\Omega_0$,
the critical radius for the collapse will depend basically on the
the energy scale of symmetry breaking. 
Generally, global textures arise after a phase transition
during the break of symmetry described by the Grand Unified Theory
(GUT), when the strong interaction is unified with the
eletroweak force. Current estimates of this unification
lead to an energy scale of $\sim 10^{16}$ GeV (Gleiser 1998).
In the same way,
if the observed fluctuations in the cosmic background radiation
are due to textures, we must have $(\delta T/T)\sim 8G\eta^2$ which implies
$\eta\sim10^{16}$ GeV in order to obtain an amplitude consistent with
the COBE measurements, $\delta T/T \sim 10^{-5}$ (e.g. Vilenkin \& Shellard 1994).

In view of these arguments, the energy scale of $10^{16}$ GeV
seems to be the natural choice of $\eta$. 
However, let's proceed here to find a new and 
independent argument for this choice.
First of all, assuming $\Omega_0=0.2$ and
$h=0.75$ $(h=H_0/100 ~{\rm km~s^{-1}~Mpc^{-1}})$,
we plot the behaviour of $R_{\rm c}$ as a function of the epoch $z_{\rm i}$
using $\eta$ as a free parameter (see Figure 1). 
Note that for earlier epochs $R_{\rm c}$
is smaller (independent of $\eta$) indicating that the first bound objects to be formed
via textures knots would have small sizes and masses. 
In fact, isocurvature fluctuations like those produced by topological
defects can evolve only after the decoupling ($z\sim10^3$), when the
Jeans mass is $\sim10^6 M_\odot$. Thus,
clumps with this mass should form first 
and evolve into larger systems through gravity.

Obviously,
the details of the evolution of clumps up to form typical
galaxies and clusters is a rather complex process, involving the thermal 
history of the gas, star formation phenomena and the mutual interaction between the clumps. It is beyond our aim to address this
problem as a whole. Instead, we proceed to simplify it by
supposing that a protogalaxy just consist of a 
regular distribution of matter (possibly
sub-galactic clumps) under a texture field
where we can apply our collapse criterion.

That being the case, 
an important quantity to be found
is the mass $M_{\rm c}$ related to a given radius $R_{\rm c}(z_{\rm i})$.
For a homegeneous universe, the non-relativistic
mass within a physical scale $\lambda$ is given by
\begin{equation}
M (\lambda) = 1.45\times 10^{11}(\Omega_0 h^2)\lambda_{\rm Mpc}^3 M_\odot
\end{equation}
(e.g. Padmanabhan 1993). This quantity is conserved during the expansion
and corresponds to the uncollapsed mass of a physical scale $\lambda$.
If we take $\lambda=R_{\rm c}$, it is possible
to study the behaviour of $M_{\rm c}$ 
as a function of the epoch $z_{\rm i}$ and the parameter $\eta$.
\begin{figure}
  \vspace{200pt}
  \caption{Plot of $R_{\rm c}$(Mpc) as a function of the redshift $z_{\rm i}$
and the energy scale of symmetry breaking $\eta$, where 
we have used $h=0.75$ and $\Omega_0=0.2$. The dotted line corresponds to the
limit.}
\end{figure}
\begin{figure}
  \vspace{200pt}
  \caption{Plot of $M$ (in solar units) as a function of the redshift $z_{\rm i}$ and the energy scale of symmetry breaking $\eta$. 
The dashed line indicates the behaviour of the Jeans mass
with $z_{\rm i}$. We have used $h=0.75$ and $\Omega_0=0.2$.}
\end{figure}
\noindent In Figure 2 we see 
that, for a same epoch $z_{\rm i}$, the value of $M$ will strongly depend on $\eta$. From the comparison between $M_{\rm c}$ and the Jeans mass,
we also note that the lowest energy scale which is able to form
self-bound systems over all the astrophysical range
($10^6-10^{15}M_\odot$) at any pos-recombination time $z_{\rm i}$ is $\eta=10^{16}$ GeV. At this energy scale, $R_{\rm c}$ is well
inside $H^{-1}/\sqrt{2}$ (see Figure 1), which defines regions where
we can consider the effects of isolated texture knots.
This result can be taken as a qualitative argument for the choice
of $\eta=10^{16}$ GeV and so we are assuming this value for the
rest of the work.

\section{Galaxy Formation}

In order to apply our collapse criterion to the
specific problem of galaxy formation, we should follow
the evolution of overdense regions containing 
typical galaxy masses ($10^{11}-10^{12}~M_\odot$). If these respect the 
condition (8), they will expand to a maximum radius, then  collapse and
eventually virialize. The dynamical properties of
the resultant systems can be estimated from the spherical model
for non-linear collapse.
Thus, at the epoch 
\begin{equation} 
(1 + z_{\rm col}) = 0.36\overline{\delta}_{\rm i}(1 + z_{\rm i}),
\end{equation}
the dissipative component of the system (i.e. baryons) reaches the virialization with the following
approximated  properties:
\begin{equation}
r_{\rm vir} = 163 (1+z_{\rm col})^{-1} \Omega_0^{-1/3} (M/10^{12}M_\odot)^{1/3}
h^{-2/3} {\rm kpc}
\end{equation}
\noindent and
\begin{equation}
\sigma_{\rm vir} = 126 (1+z_{\rm col})^{1/2} \Omega_0^{2/3} (M/10^{12}M_\odot)^{1/3}
h^{1/3} {\rm km~s^{-1}}
\end{equation}
\noindent (e.g. Padmanabhan 1993).

Before applying these expressions to 
galaxy scales, it would be interesting
to assure that the initial density contrast $\overline{\delta}_{\rm i}$
can form a self-bound system 
before the present epoch, $z_{col} > 0$. We directly find
such a condition from Eq.(18): 
\begin{equation}
\overline{\delta}_{\rm i} >  2.78(1+z_{\rm i})^{-1}.
\end{equation}

\noindent For the case of a flat model this is a simple condition
for the collapse. However, for an open universe, we 
also should have $\overline{\delta}_{\rm i} > \delta_{\rm c}$ in order to
collapse the overdense region. Using Eq.(10) and taking the limitting
case when $\delta_{\rm c} = 2.78(1+z_{\rm i})^{-1}$, we conclude that
$\Omega_0 < 0.26$ is the necessary condition for the collapse
with $z_{\rm col}>0$.

\begin{figure}
  \vspace{200pt}
  \caption{Plot $\overline{\delta}_{\rm i}$ as a function of $z_{\rm i}$.
The dotted line is the limit
$2.78(1+z_{\rm i})^{-1}$.}
\end{figure}
\begin{figure}
  \vspace{200pt}
  \caption{$\sigma-r$ plane of the virialized systems. The boxes 
delimit typical values found in the literature for spiral (hatched box) and elliptical galaxies. Curves A and B refer to $10^{12}~M_\odot$ objects
and curves C and D refer to $10^{11}~M_\odot$ objects.}
\end{figure}

In Figure 3, we plot $\overline{\delta}_{\rm i}$ as a function of $z_{\rm i}$
for initial radii that correspond to typical galaxy masses. 
The dotted line is the limit
$2.78(1+z_{\rm i})^{-1}$. Note that for $M=10^{11}M_\odot$, 
the collapse is only possible for $z_{\rm i}\la 33$ 
($z_{\rm col}\la 9$) if $\Omega_0=0.2$
and $z_{\rm i}\la 25$ ($z_{\rm col}\la 8$) if $\Omega_0=1.0$; while for $M=10^{12}M_\odot$,
$z_{\rm i}\la 15$ ($z_{\rm col}\la 5$) if $\Omega_0=0.2$ and $z_{\rm i}\la 11$ ($z_{\rm col}\la 4$)if $\Omega_0=1.0$.
These ranges in $z_{\rm i}$ produce associated loci in the 
$\sigma-r$ plane (see Figure 4). These loci show that 
our simple model texture knot$+$spherical collapse is able to
produce self-bound gravitational systems with dynamical 
properties quite similar to real galaxies. In fact, the comparison
with data compiled from the literature
shows a significant agreement between our 
predictions and typical galaxies, while it is also clear in 
Figure 4 an important trend
towards larger radii and lower velocity dispersions in our bound systems.
At the same time,
the epoch when such objects are presumed to  reach the virial equilibrium ($z_{\rm col}$) is consistent with the results
of several studies which point out an epoch of galaxy
formation by $z\la 5$ (e.g. Peebles 1993; Pahre 1998).

Now, by extending the approach developed in this section to
a wider astrophysical range ($10^6 - 10^{14}~M_\odot$),
it is possible to depict an intuitive picture of the structure
formation process when it is seeded 
by textures knots. The results
for both open and flat universes are summarized in Tables 1 and 2,
respectively, where the columns are as follows:
(1) is the mass of the systems in solar units;
(2) is the initial redshift we take;  
(3) is the initial radius of the region to be collapsed in Mpc; 
(4) is the initial average density contrast;
(5) is the redshif at which the systems reach the virialization; 
(6) is the time interval of the collapse and virialization process (the dynamical time); finally,
(7) and (8) are the radius in kpc and the velocity dispersion in km/s of
the virialized systems. 

A simple analysis of the data assembled in these tables
reveals some points we should emphasize. First of all, 
note that structure formation ceases at earlier times
in the low $\Omega_0$ universe in comparison to the flat
case. This is expected as long as matter has a smooth
distribution in high $\Omega_0$ universes so that it contributes to 
the expansiong rate but does not contribute to the gravitational
potential and so it slows down the growth of fluctuations. 
At the same time, note that the process of structure formation in our model
begins at a much earlier epoch in comparison to standard
CDM models, where most of the objects with $10^6~M_\odot$ are
formed at the epoch $z=50/b$ ($b$ is the
biasing factor) (e.g. Peebles 1993). Therefore, even 
an unbiased CDM model would present a less advanced
structure formation stage for a given redshift in
comparison to our model. 

Such an early self-bound system formation may have serious consequences.
In particular, we form the first generation of $10^6~M_\odot$ objects
at $z=135$ ($\Omega_0=0.2$) and  $z=112$ ($\Omega_0=1.0$).
These objects could be related to an early star formation
phase in the universe and its consequent reionization.
At the same time, these early spheroids may develop massive
black holes onto which matter would be accreted, possibly
forming an early generation of active galaxies.

Still concerning to the chronology of structure formation,
Gooding, Spergel \& Turok (1991) found that by $z\sim 50$, 
$\sim 3$\% of the mass
of the universe has formed non-linear objects of mass
greater than $10^6~M_\odot$ and that most of objects
larger than $10^{12}~M_\odot$ form by $z\sim 2-3$.
These results should be compared to our findings,
while in the present work we have no statistics
on the fraction of matter which is
collected in bound objects of a specific mass in a given redshift.
Note, however, that by $z\sim 50$ we already formed systems of $10^6-10^8~M_\odot$
and by $z\la3$ we form objects with $10^{13}-10^{14}~M_\odot$.
Thus, in some measure, our results are consistent with
Gooding, Spergel \& Turok (1991)

On the other hand,
a disadvantage of our model refers to the velocity dispersions of
the collapsed systems which are sistematically lower than 
those observed in astrophysical objects of same mass. 
Up to galaxy scales this effect is accompanied with
larger radii and so it can be taken as a simple consequence of the
virial theorem. On the contrary,
the lower velocity dispersion effect seems to be
particularly serious in larger scales, like galaxy clusters,
where the model predicts objects with radii similar to those found
in real clusters. Actually,
this is the opposite one could expect in a texture
scenario, where the positive skewness of the mass fluctuations
induces higher velocity dispersions for objects of a given
mass than (as an example) in standard CDM models
(e.g. Bartlett, Gooding \& Spergel 1993). We think this trend
is probably due to the fact we are considering only
the dissipative baryonic component of matter inside $r_{\rm vir}$.
By adding dark matter within this radius
we could recover values of velocity dispersion nearer
to reality. For instance, in the case 
of a galaxy cluster scale ($10^{14}~M_{\odot}$), if the 
total mass (dissipative plus non-dissipative
components) inside $r_{\rm vir}$ reached $10^{15}~M_{\odot}$ the
velocity dispersion of the collapsed system would be 
583 km/s ($\Omega_0=0.2$) or 1475 km/s ($\Omega_0=1.0$), which
could be taken as more ``normal" values. However, if this is the
right solution to the problem,  we would have
to assume specific distributions for
the luminous and dark matter components when considering different objects.
At the same time, we cannot
discard the possibility that the simplicity of our model texture 
knot$+$spherical collapse is not taking account all of the
physics relevant to the structure formation process. Indeed, we are ignoring
interactions between clumps, star formation and hydrodynamical effects.
These physical mechanisms are probably important over the systems evolution 
and their inclusion would change (to some extent)
the outcomes of the model.

\begin{table}
 \caption{Properties of self-bound gravitational systems
seeded by textures knots in a $\Omega_0=0.2$ universe.}
\label{symbols}
\begin{tabular}{@{}lccccccc}
$M$   & $z_{\rm i}$ & $r_{\rm i}$ & $\overline{\delta}_{\rm i}$ &$z_{\rm col}$ & $t_{\rm col}$ & $r_{\rm vir}$ & $\sigma_{\rm vir}$ \\
$10^6$    & 1000 & 0.04 & 0.0067 & 135 & $7.7\times 10^6$ & 1.4 & 4.6 \\
$10^7$    & 700  & 0.09 & 0.0038 & 80  & $1.6\times 10^7$ & 7.3 & 7.5 \\
$10^8$    & 350  & 0.18 & 0.0076 & 50  & $3.3\times 10^7$ & 16  & 13  \\
$10^9$    & 150  & 0.39 & 0.0205 & 27  & $7.6\times 10^7$ & 34  & 21  \\
$10^{10}$ & 70   & 0.85 & 0.0416 & 16  & $1.5\times 10^8$ & 73  & 35  \\
$10^{11}$ & 32   & 1.84 & 0.0886 & 9   & $3.0\times 10^8$ & 155 & 58  \\
$10^{12}$ & 14   & 3.87 & 0.2131 & 5   & $5.4\times 10^8$ & 333 & 98  \\
$10^{13}$ & 6    & 8.57 & 0.4273 & 2.7 & $8.3\times 10^8$ & 680 & 163 \\
$10^{14}$ & 2.4  & 18.4 & 0.8100 & 1.2 & $1.2\times 10^9$ & 1465& 271 \\
\end{tabular}
\end{table}

\begin{table}
 \caption{Properties of self-bound gravitational systems
seeded by textures knots in a $\Omega_0=1.0$ universe.}
\label{symbols}
\begin{tabular}{@{}lccccccc}
$M$   & $z_{\rm i}$ & $r_{\rm i}$ & $\overline{\delta}_{\rm i}$ &$z_{\rm col}$ & $t_{\rm col}$ & $r_{\rm vir}$ & $\sigma_{\rm vir}$\\
$10^6$    & 1000 & 0.02 & 0.0054 & 112 & $6.9\times 10^6$ & 1.4  & 12 \\
$10^7$    & 550  & 0.05 & 0.0052 & 66  & $1.5\times 10^7$ & 4.3  & 20 \\
$10^8$    & 260  & 0.11 & 0.0100 & 39  & $3.2\times 10^7$ & 9.2  & 34 \\
$10^9$    & 115  & 0.23 & 0.0261 & 23  & $6.7\times 10^7$ & 20   & 56 \\
$10^{10}$ & 55   & 0.50 & 0.0491 & 13  & $1.4\times 10^8$ & 42   & 93 \\
$10^{11}$ & 25   & 1.07 & 0.1070 & 8   & $2.5\times 10^8$ & 91   & 157 \\
$10^{12}$ & 11   & 2.31 & 0.2330 & 4   & $5.7\times 10^8$ & 190  & 261 \\
$10^{13}$ & 5    & 4.98 & 0.4020 & 2   & $1.1\times 10^9$ & 425  & 422 \\
$10^{14}$ & 1.6  & 10.7 & 1.0000 & 0.7 & $8.5\times 10^8$ & 857  & 685 \\
\end{tabular}
\end{table}

\section{Discussion and Summary}

In this work, we developed a quantitative criterion for the
collapse of a spherical distribution of matter under a single texture
field. Further, we applied this criterion to form galaxies in
an idealized smooth universe where texture knots can be taken
as isolated sources for gravitational collapse.
Despite the simplicity of the model, the properties
of the systems which collapse and virialize respecting
our criterion are similar to real
astrophysical objects. A drawback of the model
is the trend towards larger radii and lower velocity dispersions
in our bound systems in comparison to typical values found
in the literature for real objects of same mass. This trend
is specifically significant for velocity dispersions and could be
related to the quantity of dark matter inside the virialization
radius. However, we would like to emphasize that the most important 
achievement in this work is the determination of a collapse criterion 
using simple arguments and the possibility of
applying it in a direct way to predict astrophysical
objects. This means that local effects of isolated knots
may be important to the development of structures in the universe
and suggests that individual textures (not only their dynamical collective
effect) should be taken account in galaxy formation models.

In the following, we summarize the main results of this work:

\begin{enumerate}
\renewcommand{\theenumi}{(\arabic{enumi})}

\item We showed that, under specific conditions,
the destiny of an overdense
region under a single 
texture field will basically depend on its initial size
and the energy scale of symmetry breaking $\eta$ (if $\Omega_0 < 1$).
For those regions smaller than
$R_{\rm c}$, the system will expand to a maximum radius and
then inevitably collapse. 

\item We found an additional argument for the
choice of $\eta$ as $10^{16}$ GeV. This is simply the
lowest energy scale which allows the
formation of structures on scales relevant to astrophysical objects
($10^6-10^{15}M_\odot$) at any time $z_{\rm i}<1000$.

\item We showed 
the loci of objects with masses $10^{11}-10^{12}M_\odot$
in the $\sigma-r$ plane agree well with the observed properties of real galaxies. 

\item Our results also indicate
an epoch of galaxy formation fairly consistent with
other independent studies which point out it should
have been by $z\sim 5$.

\item Our model produces a chronology of structure formation
which begins at earlier times in comparison to other models,
in particular CDM models.

\item The existence of spheroids by $z\sim 100$
suggests the possibility of associated phenomena like early star formation process, reionization
of the universe, production of black holes and active galaxies.

\end{enumerate}

\section*{Acknowledgments}
We thank the referee A. Liddle for useful
suggestions. We also thank R. de Carvalho and
R. Coziol for the critical reading of the paper.
Finally, we thank the finantial support of
the Brazilian FAPESP and the CNPq.

\bsp

\label{lastpage}

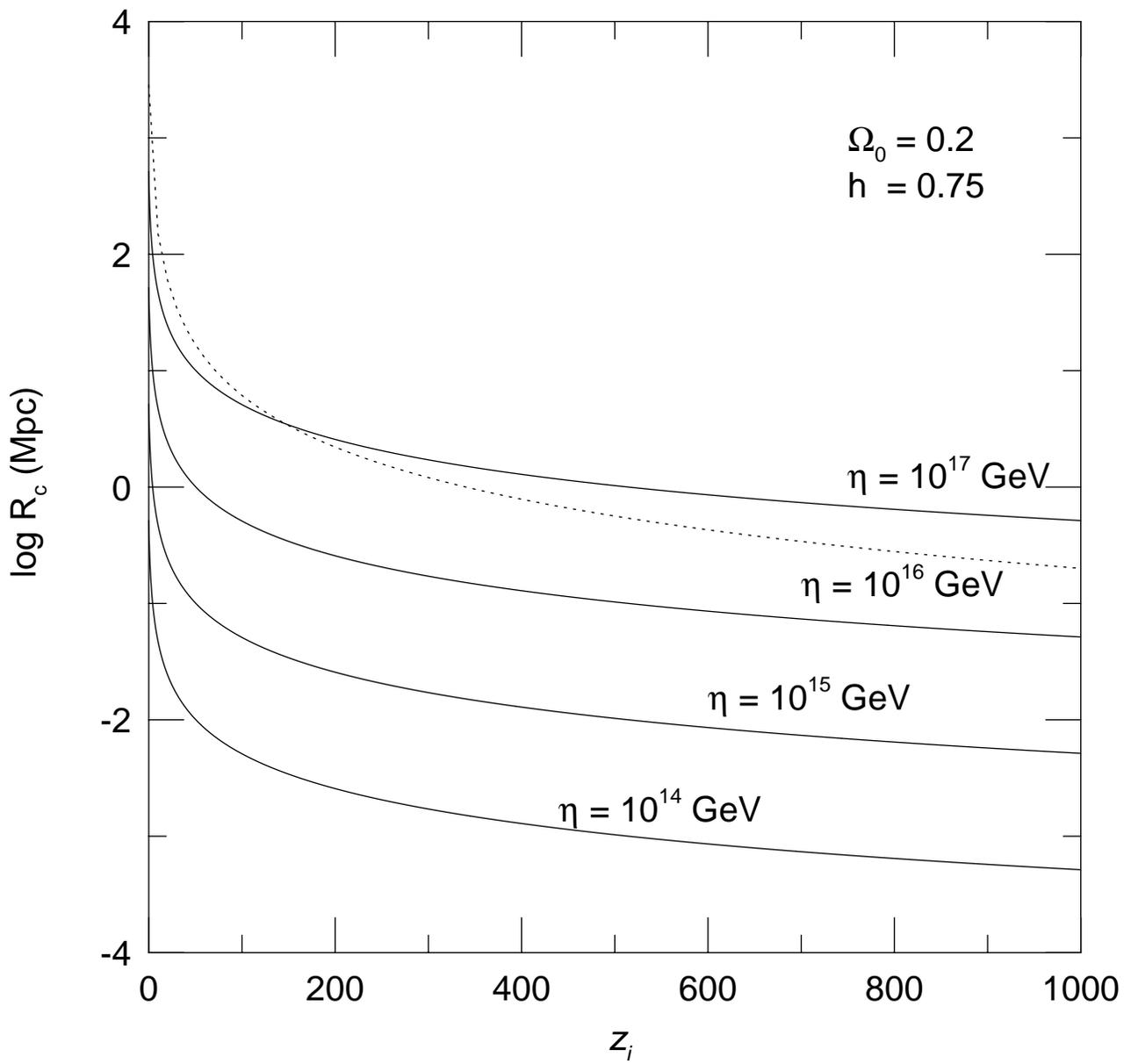
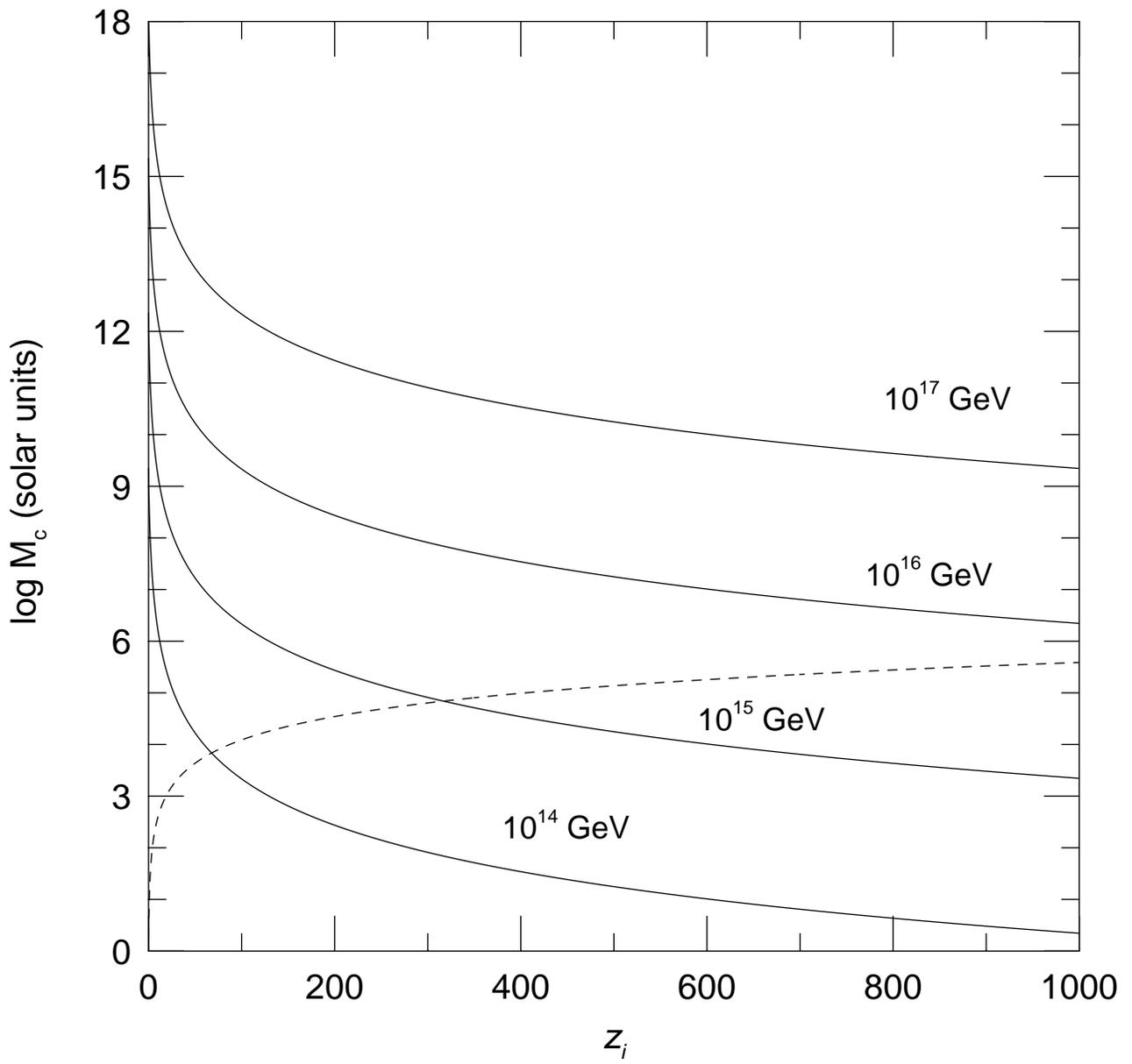
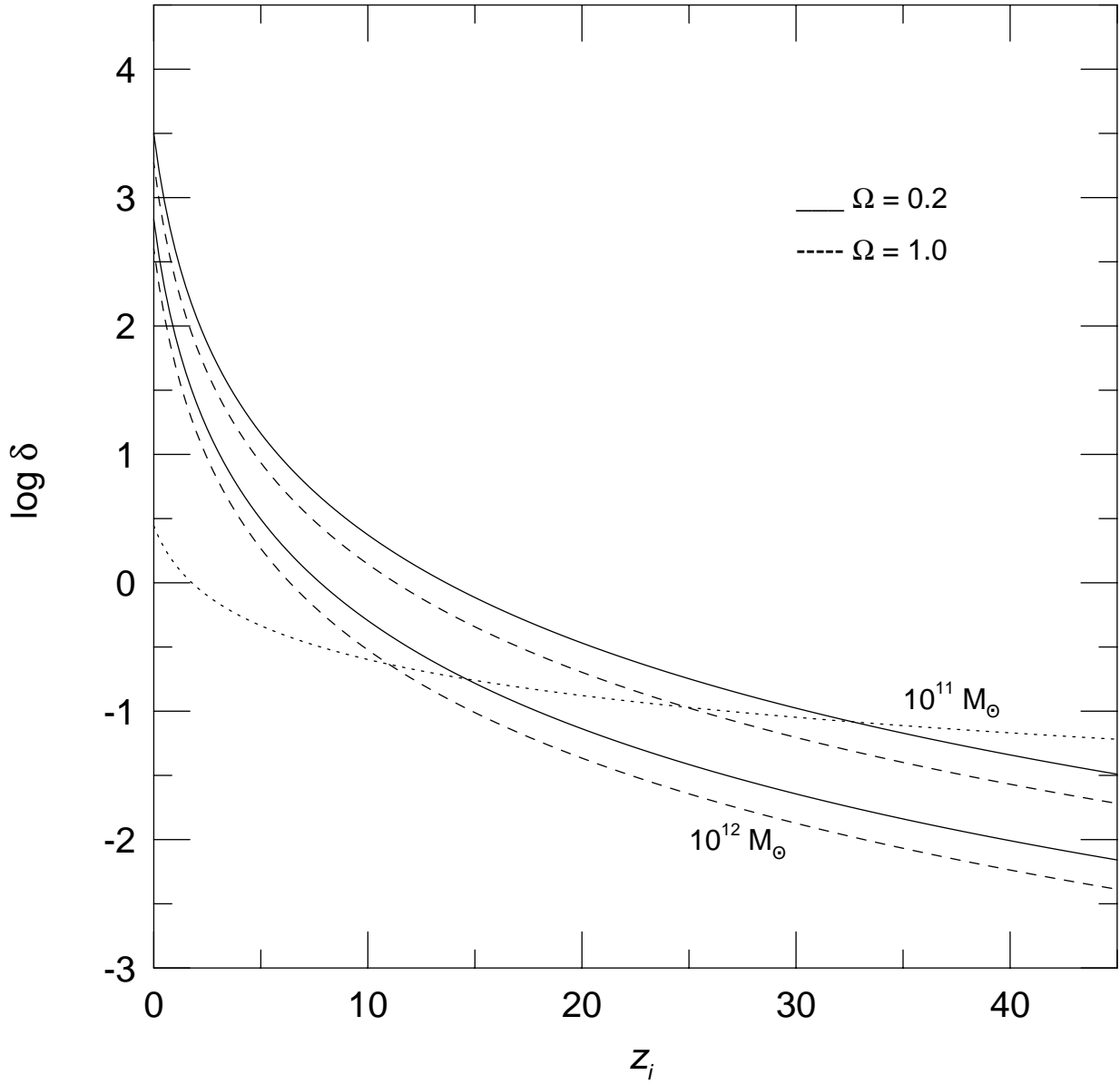
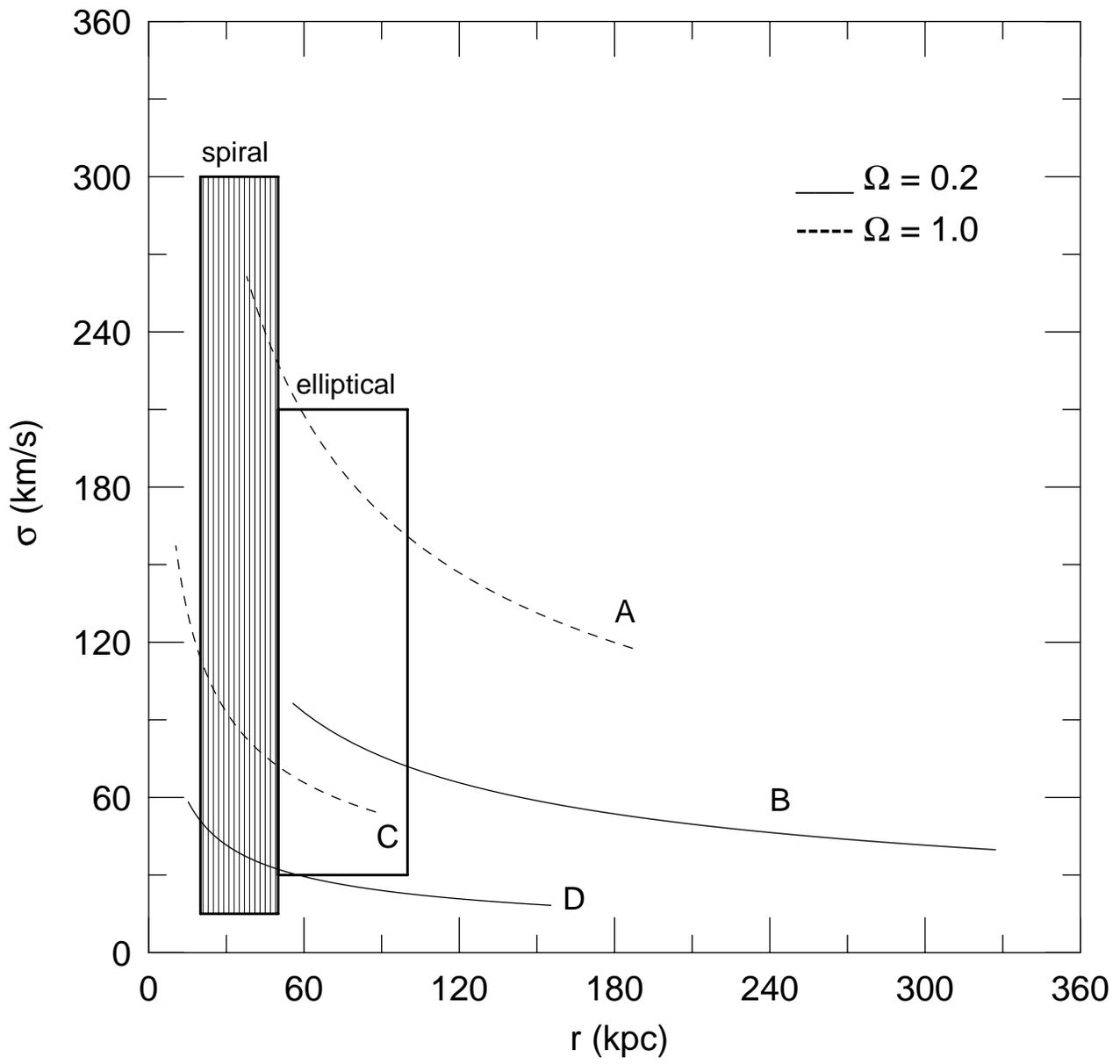
\end{document}